\documentclass[%
 reprint,
superscriptaddress,
 amsmath,amssymb,
 aps,
prl,
]{revtex4-2}

\usepackage{float}
\usepackage{physics}
\usepackage{graphicx}
\usepackage{dcolumn}
\usepackage{bm}
\usepackage{xcolor}

\usepackage{xr}
\usepackage{subfigure}
\makeatletter

\newcommand*{\addFileDependency}[1]{
\typeout{(#1)}
\@addtofilelist{#1}
\IfFileExists{#1}{}{\typeout{No file #1.}}
}\makeatother

\newcommand*{\myexternaldocument}[1]{%
\externaldocument{#1}%
\addFileDependency{#1.tex}%
\addFileDependency{#1.aux}%
}
\myexternaldocument{main_supplementary_material}

\newcommand{\expV}[2]{\langle #1 \rangle_{#2}}
\newcommand{\varV}[2]{\Delta #1^2_{#2}}
\newcommand{\stdV}[2]{\Delta #1_{#2}}

\newcommand{\disp}{\text{disp}}
\newcommand{\TTR}{\text{TTR}}
\newcommand{\td}{\text{d}}
\newcommand{\crit}{\text{crit}}
\newcommand{\tss}{\text{ss}}
\newcommand{\T}{\text{T}}
\newcommand{\R}{\text{R}}
\newcommand{\TR}{\text{TR}}
\newcommand{\g}{\text{g}}
\newcommand{\C}{\text{C}}
\newcommand{\J}{\text{J}}

\begin{document}

\title{Ancilla-Assisted Accelerated Readout of Transmon Qubits}

\author{Christopher Corlett}
\affiliation{Quantum Engineering Centre for Doctoral Training, University of Bristol, Tyndall Avenue, Bristol, BS8 1FD, UK}
\affiliation{H. H. Wills Physics Laboratory, University of Bristol, Tyndall Avenue, Bristol, BS8 1TL, UK}

\author{Noah Linden}
\affiliation{School of Mathematics, University of Bristol, Fry Building, Woodland Road, Bristol, BS8 1UG, UK}

\author{Paul Skrzypczyk}
\affiliation{H. H. Wills Physics Laboratory, University of Bristol, Tyndall Avenue, Bristol, BS8 1TL, UK}

\date{\today}

\begin{abstract}
We present a scheme for accelerating dispersive readout of superconducting transmon qubits. 
Dispersive readout works by weakly coupling the target transmon to a resonator, such that the resonator experiences a state-dependent frequency shift, which can then be read out. Our  scheme  introduces an ancillary transmon coupled to the same resonator. By appropriately correlating its state with that of the target transmon (using a fast CNOT gate), we show that the resonator frequency shift is approximately doubled, halving the required measurement duration to maintain the same quality.  We present analytical results within the dispersive approximation, before employing and extending the recently introduced technique of branch analysis to account for ionization and other nonlinearities. Using this, and taking into account realistic noise and gate times, we show that our scheme allows measurement durations smaller by a factor of approximately 0.6, demonstrating that it offers a viable route for accelerating measurements of transmon qubits. 
\end{abstract}

\maketitle

\subsection{Introduction}
The ability to perform high quality measurements is vital to all quantum technologies. However, in many quantum computing architectures, high quality measurements are slow compared to the duration of other operations. This is the case for superconducting transmon qubits, the focus of this Letter. For example, mid-circuit measurements used during syndrome extraction are a significant bottleneck in the total execution time of fault tolerant algorithms \cite{harper2025characterisingfailuremechanismserrorcorrected}.

In this Letter, we will introduce a scheme for maintaining high quality while  speeding up measurement of the widely used \textit{transmon} qubit \cite{Koch_2007, Blais_2021, Krantz_2019}.

Transmons are typically measured using a technique known as \textit{dispersive readout} \cite{Blais_2021, Krantz_2019, Blais_2004, Gambetta_2008}. When capacitively coupled in the \textit{dispersive regime} -- wherein the transmon-resonator (TR) detuning $\Delta$ is large compared to the coupling $g$ -- the frequency of the resonator shifts dependent upon the state of the transmon by an amount known as the \textit{dispersive shift} $\chi\approx \frac{g^2}{\Delta}$. This shift is then determined by driving the resonator and measuring the reflected signal. As a consequence, the transmon state is measured.   

The standard measure of quality for dispersive measurement is the \textit{signal to noise ratio} (SNR). After making the dispersive approximation, it can be shown that the optimal SNR in the steady state regime is \cite{Gambetta_2008, Blais_2021}
\begin{equation}\label{eqn:snr1_ss}
    S(\tau)\approx 4\sqrt{\eta n_{\tss}|\chi| \tau}, 
\end{equation}
where $\tau$ is the measurement duration, $\eta$ the detector efficiency and $n_{\tss}$ the mean photon number in the resonator steady state.

While dispersive readout is very effective, to achieve a similar quality to the best gates, it must operate orders of magnitude slower than the slowest gate speed \cite{harper2025characterisingfailuremechanismserrorcorrected, google_2024, Bengtsson_2024}. This is primarily because the TR system must remain within the dispersive regime during the measurement process. This requires both that $\Delta \gg g$, and that the system remains in a low-energy state throughout the process. A faster measurement could be achieved in principle by increasing $\chi$ (increasing $g$ or decreasing $\Delta$) or increasing $n_{\tss}$ (by increasing the resonator drive strength). However, this generally takes the system out of the dispersive regime, leading to non-linear effects which reduce the quality of the measurement, and deviations from \eqref{eqn:snr1_ss}.

One of the most detrimental non-linear effects is transmon \textit{ionization}; if the resonator is driven too strongly then the transmon can be excited into a non-computational state \cite{Sank_2016, Khezri_2023, xiao2024diagrammaticmethodcomputeeffective, Shillito_2022, Dumas_2024, Hoyau_2026}. The number of photons in the resonator above which ionization is expected to occur is known as the \textit{critical photon number} $n_{\crit}$. Crucially, $n_{\crit}$ decreases with smaller $\Delta$ and larger $g$. Ionization thus limits the driving strength of the resonator; $n_{\tss}$ must be kept suitably below $n_{\crit}$. A key challenge in optimising dispersive readout is to operate the system at the limit of the dispersive regime, with $n_{\tss}$ and $\chi$ as large as possible while keeping ionization negligible.

Recently, there have been a number of new strategies for faster and higher-quality readout. New circuit designs with alternative TR couplings have been investigated \cite{Chapple_2025, Dassonneville_2020, Beaulieu_2026, Mori_2025, Chapple_2025_a, Ye_2024, Salunkhe_2025}, overcoming the perturbative coupling used in dispersive readout, as well as improvements from considering the TR detuning \cite{Swiadek_2024, Kurilovich_2025}. There is still however ample room for alternative approaches to accelerating dispersive readout.

\begin{figure}
    \centering
    \includegraphics[width=0.7\linewidth]{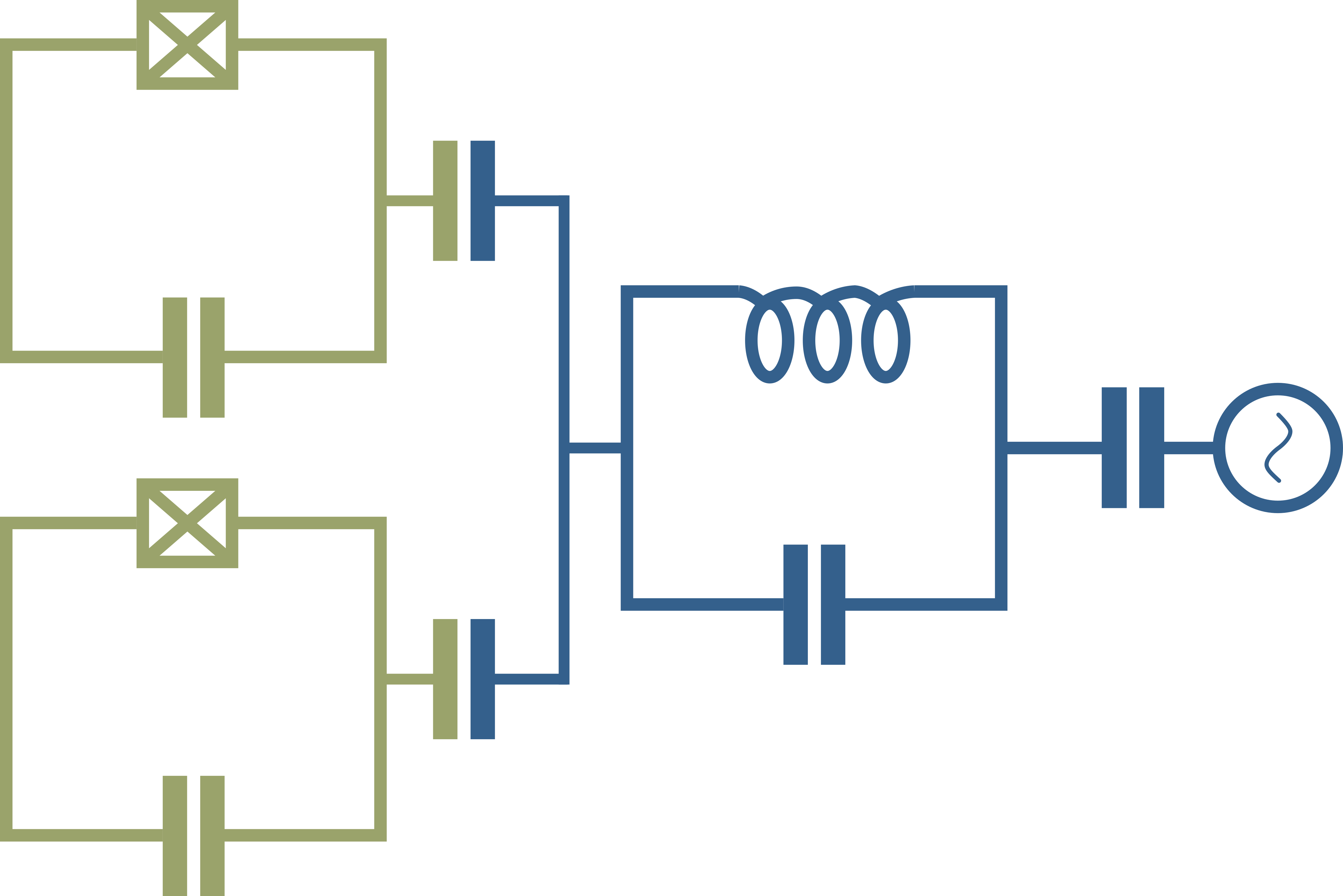}
    \caption{\textbf{Ancilla-assisted dispersive readout scheme.} Two transmons (green) capacitively coupled to a driven resonator (blue).}
    \label{fig:scheme_transmons}
\end{figure}

In this Letter, we propose a novel approach from a different direction. We maintain dispersive coupling, but instead achieve speed-up by introducing an ancillary transmon. We show that by coupling this additional transmon to the resonator, as in Fig. \ref{fig:scheme_transmons}, the total dispersive shift can be approximately doubled. This is achieved by correlating the state of the ancillary transmon to the original transmon by the action of a (fast) CNOT gate. Performing dispersive readout via the resonator now results in a faster measurement, without sacrificing quality.

Our approach takes inspiration from our previous scheme for space-time trade-offs in quantum measurements \cite{Corlett_2025}. A direct application of this scheme requires an additional resonator. A key contribution here is to tailor the scheme to dispersive readout, allowing us to achieve the same speed-up with only a single resonator.

\begin{figure}[!t]
    \centering
    \includegraphics[width=\linewidth]{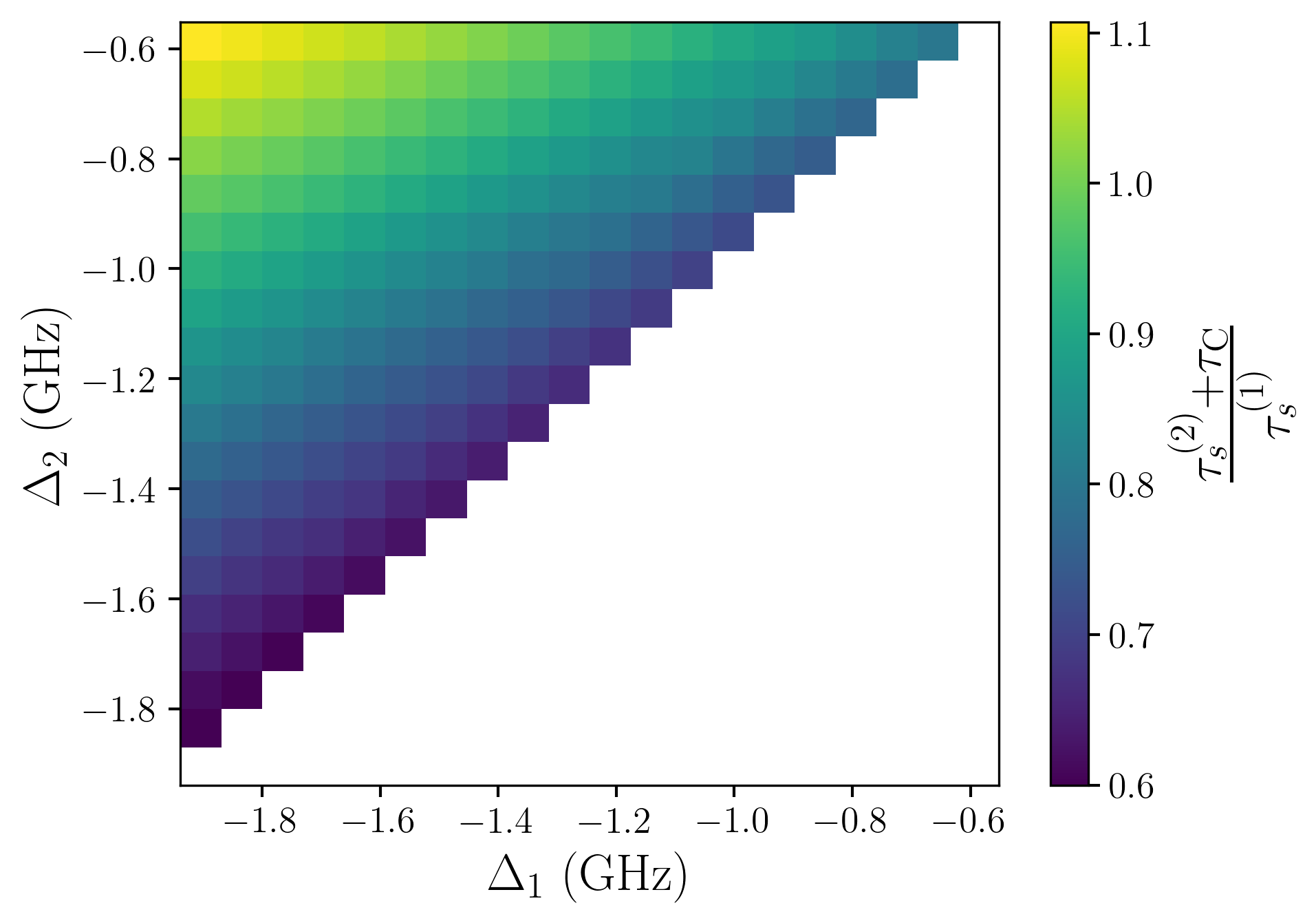}
    \caption{\textbf{Space-time trade-off in dispersive readout.} Heat map of duration ratio $\frac{\tau^{(2)}_6 + \tau_\text{C}}{\tau^{(1)}_6}$. Here, $\tau^{(2)}_6$ is the duration for the TTR system SNR \eqref{eqn:snr_noise} to reach a value of 6, $S_p^{(2)}(\tau^{(2)}_6)=6$ ($\tau^{(1)}_6$ is the TR system equivalent), and $\tau_\text{C}$ is the duration of a CNOT gate. Darker values (closer to equal detuning) correspond to faster speed-up, which is achieved in most of the parameter space. System parameters \eqref{eqn:htr} are $E_{\J,1}=E_{\J,2}=24.2\,\text{GHz}$, $n_{\g,1}=n_{\g,2}=0$, $g_1=g_2=0.12\,\text{GHz}$ and $\omega_\R=7.5\,\text{GHz}$. $E_{\C,1}, E_{\C,2}\in[0.17,0.27]\,\text{GHz}$ vary across the horizontal and vertical axis respectively, shown as the corresponding change in $\Delta_1$ and $\Delta_2$. We take $\eta = 0.5$, CZ error rate $p=0.0026$, and $\tau_\text{C}=55\,\text{ns}$ (duration of CZ + Hadamard. The first Hadamard of the CNOT decomposition can be absorbed into the preparation of the ancilla), from Google's Willow processor \cite{google_2024}. Parameters for the standard TR system readout are the same as those for the TTR system except for $E_{\C} = \max\{E_{\C,1},E_{\C,2}\}$, chosen to minimise $\tau^{(1)}_6$  (comparing the TTR system against the faster of the two TR systems). Finally, we take $n_\tss = 18.$
    }
    \label{fig:snr_s_ratios}
\end{figure}

As we will show below in detail, the optimal SNR under the dispersive approximation for our scheme (using superscripts to denote the number of transmons coupled to the resonator from hereon) is
\begin{equation}\label{eqn:snr2_ss}
    S^{(2)}(\tau^{(2)})\approx4\sqrt{\eta n_{\tss}^{(2)}|\chi_1+\chi_2| \tau^{(2)}},    
\end{equation}
where $\chi_1$ and $\chi_2$ are the dispersive shifts due to the target and ancillary transmons respectively, which throughout we take to be negative (arising from negative detunings).

Comparing \eqref{eqn:snr2_ss} against \eqref{eqn:snr1_ss} (with appropriate superscripts added), under the condition of equal resonator steady-state photon number $n_{\tss}^{(2)}=n_{\tss}^{(1)}$ (and efficiency $\eta$),  then $S^{(2)}\left(\left|\frac{\chi_1}{\chi_1 + \chi_2}\right|\tau^{(1)}\right)=S^{(1)}(\tau^{(1)})$. A `space-time trade-off' is thus achieved: using our new transmon-transmon-resonator (TTR) readout scheme, a measurement of shorter duration
\begin{equation}\label{eqn:space-time-trade-off}
    \tau^{(2)}=\left|\frac{\chi_1}{\chi_1 + \chi_2}\right|\tau^{(1)},
\end{equation}
 has a quality equal to a slower measurement, of duration $\tau^{(1)}$, using the standard TR readout.

The key technical challenge that remains in showing our scheme works is ensuring the previously discussed non-linear effects are not significantly amplified by coupling a second transmon to the resonator, which could invalidate the speed-up. To this end, we must ensure that the critical photon number for the double transmon device $n_{\crit}^{(2)}$ is not significantly less than $n_{\crit}^{(1)}$, such that it is reasonable to set $n_{\tss}^{(2)}\approx n_{\tss}^{(1)}$. We must also consider the non-linear dependence of the resonator photon number on the system parameters, which were neglected in \eqref{eqn:snr1_ss} and \eqref{eqn:snr2_ss}. This results in a photon-number-dependent dispersive shift $\chi_{12}(n)$ (where $\chi_{12}(0)\approx\chi_1+\chi_2$ and $n$ is the photon number), which becomes relevant at higher drive powers \cite{Boissonneault_2008, Boissonneault_2009, Boissonneault_2010}. 

Finally, a CNOT gate is integral to our scheme. We envisage this will be facilitated via a fast tunable coupler, such as used in Google's Willow processor \cite{google_2024}. This is an effective way to mitigate ionization processes induced by transmon-transmon coupling \cite{Hoyau_2026} and, hence, we do not include a direct transmon-transmon coupling in our model. We will need to model realistic noise in gate application, as well as the finite duration of the gate, to ensure neither significantly reduces the speed-up achieved by our scheme.

In the remainder of this Letter we will show that none of the above concerns significantly reduce the speed-up achieved by our scheme. We begin with the simplified analysis under the dispersive approximation, before moving on to the general analysis, taking into account non-linear effects and realistic CNOT gates. The final results are summarised in Fig.~\ref{fig:snr_s_ratios}, where we demonstrate that under the general analysis, we achieve a substantive acceleration for a wide range of TR detunings.

\subsection{Dispersive Regime Analysis}
The Hamiltonian for the transmon-transmon-resonator (TTR) system is a simple extension of the Hamiltonian for the TR system \cite{Blais_2021, Krantz_2019, Lalumiere_2010, Filipp_2009, Kubica_2023, Hung_2026} (details in SM A) ($\hbar=1$): 
\begin{multline}\label{eqn:htr}
    H_{\TTR} = \sum_{q=1,2}\left(4E_{\C,q}\left(n_{\T,q}-n_{\g,q}\right)^2 - E_{\J,q}\text{cos}\left(\phi_{\T,q}\right)\right) \\
    + \omega_{\R}a^\dagger a - \sum_{q=1,2}\left(ig_{q} \left(n_{\T,q}-n_{\g,q}\right) (a-a^\dagger)\right),
\end{multline}
where $E_{\C,q}$ is the charging energy, $E_{\J,q}$ the Josephson energy, $n_{\g,q}$ the gate charge, $n_{\T,q}$ and $\phi_{\T,q}$ the transmon charge and phase conjugate operators, $\omega_{\R}$ the resonator frequency, $a$ the resonator annihilation operator and $g_{q}$ the coupling strength. The subscript $q\in\{1,2\}$ specifies which transmon. The system is operated in the transmon regime $E_{\J,q}/E_{\C,q}\gg1$ which suppresses noise on the qubit subspace from the gate charge $n_{\g,q}$ \cite{Koch_2007}.

The resonator is driven by the time-dependent term 
\begin{equation}\label{eqn:drive}
    H_{\td}(t) = -i2\epsilon_{\td}(t)\text{sin}(\omega_{\td}t)(a - a^\dagger),
\end{equation}
where $\epsilon_{\td}(t)$ is the drive amplitude and $\omega_{\td}$ the drive frequency. It will also experience decay at a rate $\kappa$.

Driving the resonator and measuring the reflected signal acts as an effective measurement of the transmons. To understand this: (i) the Schrieffer-Wolff transformation (SWT) \cite{Blais_2021, Krantz_2019, Blais_2004} is performed, which approximately diagonalises the Hamiltonian \eqref{eqn:htr} in the dispersive regime $\frac{g_{q}}{|\Delta_{q}|}\ll1$, where $\Delta_{q}=\omega_{01,q}-\omega_\R$ are detunings, with $\omega_{01,q}=E_{1,q}-E_{0,q}$, and $E_{k,q}$ the energy levels of each transmon. (ii) The dispersive approximation is made, which truncates the transmon Hilbert spaces to qubit subspaces and neglects terms from the SWT considered negligible in the dispersive regime: the off-diagonal terms of order $g_{q}^2/\Delta_{q}$, which are fast rotating and removed by the rotating wave approximation, and all terms of order $g_{q}^3/\Delta_{q}^2$ or higher. The Hamiltonian becomes (in the SWT basis)
\begin{equation}\label{eqn:H_disp}
    H_{\disp}=\sum_{q=1,2}\frac{\omega_{\T,q}}{2}\sigma_{Z,q} + \left(\tilde{\omega}_{\R} + \sum_{q=1,2}\chi_q\sigma_{Z,q}\right)a^\dagger a,
\end{equation}
where $\omega_{\T,q}$ is the frequency of the transmon's qubit subspace, $\tilde{\omega}_{\R}$ the dressed resonator frequency, $\chi_q$ the dispersive shift and $\sigma_{Z,q}=\ketbra{1}{1}-\ketbra{0}{0}$. 

The change to the drive term is negligible under the dispersive approximation and, as long as a Purcell filter is used, the decay term is also unchanged \cite{Blais_2021, Krantz_2019, Sete_2015}. Including the resonator drive and decay, the evolution of the system is given by
\begin{equation}\label{eqn:lindblad_disp}
    \frac{\partial}{\partial t}\rho = -i[H_{\disp} + H_{\td}(t),\rho] + \kappa\mathcal{D}[a]\rho,
\end{equation}
where $\rho$ is the density operator of the combined system in the truncated SWT basis and $\mathcal{D}[a]\rho = a\rho a^\dagger - \{a^\dagger a, \rho\}/2$ is the Lindblad dissipator. The full derivation of \eqref{eqn:lindblad_disp} and its solution can be found in Supplemental Material (SM) B and C respectively, closely following \cite{Lalumiere_2010, Blais_2007}.

The transmons are measured by probing the resonator frequency using homodyne measurement, performed by driving the resonator at its dressed frequency $\tilde{\omega}_\R$ and mixing the reflected signal with a local oscillator (LO) at identical frequency $\tilde{\omega}_\R$. The term $(\chi_1\sigma_{Z,1}+\chi_2\sigma_{Z,2})a^\dagger a$ in \eqref{eqn:H_disp} couples the transmon states and the resonator frequency, shifting the resonator frequency to $\tilde{\omega}_\R~-~(-1)^{j_1}\chi_1-(-1)^{j_2}\chi_2$, dependent on the state of the transmons $\ket{j_1j_2}$, for $j_1$, $j_2 \in \{0, 1\}$. This in turn causes the phase of the reflected signal to shift. By measuring the phase difference between the LO and the reflected signal this shift is determined \cite{Blais_2021, Krantz_2019}. 

Homodyne measurement outputs an integrated voltage signal $V_{\ket{j_1j_2},\tau}$, where $\tau$ is the measurement duration. $V_{\ket{j_1j_2},\tau}$ is a random variable with distribution $\mathcal{P}_{\ket{j_1j_2},\tau}(v)$. By inferring which distribution $V_{\ket{j_1j_2},\tau}$ came from, using maximum likelihood estimation, the state of the transmons is measured \cite{Lalumiere_2010, Blais_2007, Filipp_2009}. 

In our scheme we speed-up the measurement of the target transmon (in the state $\ket{j_1}\in\{\ket{0},\ket{1}\}$) by first applying a CNOT gate from  target to ancillary transmon (initialised in $\ket{0}$), resulting in the state $\ket{j_1j_1}$. We therefore must distinguish between $\mathcal{P}_{\ket{00},\tau}(v)$ and $\mathcal{P}_{\ket{11},\tau}(v)$.

The distinguishability between two distributions $\mathcal{P}_1$ and $\mathcal{P}_2$ is quantified by the SNR \cite{Blais_2021, Didier_2015, Gambetta_2008,  Chapple_2025} \footnote{Note these sources have slightly different definitions of the SNR but results derived are consistent. We use a definition consistent with \cite{Chapple_2025, Gambetta_2008} which is a factor of $\sqrt{2}$ larger than the definition given in \cite{Blais_2021, Didier_2015}.}
\begin{equation}\label{eqn:snr}
    S\left(\mathcal{P}_1, \mathcal{P}_2\right) = \frac{2\left|\expV{V}{\mathcal{P}_1}-\expV{V}{\mathcal{P}_2}\right|}{\stdV{V}{\mathcal{P}_1} + \stdV{V}{\mathcal{P}_2}},
\end{equation}
where $\expV{V}{\mathcal{P}_\ell}$ is the mean voltage and $\stdV{V}{\mathcal{P}_\ell}$ the standard deviation for distribution $\mathcal{P}_\ell$. The form of ${\mathcal{P}_{\ket{j_1j_2},\tau}(v)}$ is complicated in the transient regime, but in the steady-state regime it is approximately Gaussian \cite{Landi_2024}. Applying the theory of continuous measurement \cite{Srinivas_1981, Wiseman_1993, Landi_2024} to \eqref{eqn:lindblad_disp}, the SNR $S\left(\mathcal{P}_{\ket{00},\tau}, \mathcal{P}_{\ket{11},\tau}\right)$, which we denote by $S^{(2)}(\tau)$, equals
\begin{align}\label{eqn:snr_continuous_measurement}
     S^{(2)}(\tau) = 2\sqrt{\frac{\eta\kappa}{\tau}}\left|\int_0^\tau\text{Re}[(\alpha_{00}(t)-\alpha_{11}(t))]dt\right|,
\end{align}
where $\alpha_{j_1j_1}(t)$ are (time-dependent) parameters of the resonator coherent state, $\ket{\alpha_{j_1j_1}(t)}~=~e^{-|\alpha_{j_1j_1}(t)|^2/2}\sum_{n=0}^\infty\frac{(\alpha_{j_1j_1}(t))^n}{\sqrt{n!}}\ket{n}$, which are solutions to
\begin{multline}\label{eqn:coherent_state_solution}
    \frac{\partial}{\partial t}\alpha_{j_1j_1}(t) = 2\epsilon_\td(t)\text{sin}(\omega_\td t)e^{i\omega_\td t} -i[(\tilde{\omega}_\R - \omega_\td)  \\
    -(-1)^{j_1}(\chi_1+\chi_2)]\alpha_{j_1j_1}(t) -\frac{\kappa}{2}\alpha_{j_1j_1}(t).
\end{multline}
A derivation and analytical solution to \eqref{eqn:snr_continuous_measurement} is provided in SM D. We also show how \eqref{eqn:snr2_ss} is recovered in the steady state regime when using an optimal choice of parameters: $\kappa^*=2|\chi_1+\chi_2|$, $\omega^*_\td=\tilde{\omega}_\R$ and $\epsilon^*_\td(t)=\sqrt{2n_{\tss}^{(2)} |\chi_1+\chi_2|^2}$ (independent of $t$), and using the fact that, for $t \gg 2/\kappa$,  $n_{\tss}^{(2)}~=~|\alpha_{00}(t)|^2 = |\alpha_{11}(t)|^2$. 

Thus, we have shown that, under the dispersive approximation, our scheme achieves a readout speed-up, as described in \eqref{eqn:space-time-trade-off}. 

\subsection{Beyond Dispersive Regime Analysis}
To show that our scheme continues to work under realistic conditions we now consider the effect of non-linear terms and realistic CNOT gates, which were neglected in the above.

The dispersive shift $\chi_1+\chi_2$ and the dressed frequency $\tilde{\omega}_\R$ in \eqref{eqn:coherent_state_solution} were derived under the dispersive approximation, neglecting higher order non-linear terms \cite{Boissonneault_2008, Boissonneault_2009, Boissonneault_2010}. With higher drive power these terms become relevant, even at photon number populations below the critical photon number \cite{Gambetta_2006, Schuster_2005}, so the effective dispersive shift and dressed resonator frequency will drift as the resonator is driven. The CNOT gate will also be noisy and takes finite duration. In order to make a more complete comparison between our scheme and the standard readout method we now consider the effect of the above.

We follow \cite{Shillito_2022} and introduce the photon-number-dependent resonator frequency $\omega_{\R,j_1j_2}(n(t))$, dependent on the transmon state $\ket{j_1j_2}$ and the mean photon number $n(t)=|\alpha_{j_1j_2}(t)|^2$ in the resonator. In what follows, we drop the time dependence $t$ on $n(t)$ for brevity. The photon number dependent dispersive shift and dressed resonator frequency are $\chi_{12}(n)~=~\frac{1}{2}(\omega_{\R,11}(n)~-~\omega_{\R,00}(n))$ and $\tilde{\omega}_\R(n)~=~\frac{1}{2}(\omega_{\R,11}(n)~+~\omega_{\R,00}(n))$ respectively. In the low energy regime they satisfy $\chi_{12}(0)~\approx~\chi_1 + \chi_2$ and $\tilde{\omega}_\R(0)\approx\tilde{\omega}_\R$. Using $\omega_{\R,j_1j_2}(n)$, the coherent state dynamics \eqref{eqn:coherent_state_solution} becomes
\begin{multline}\label{eqn:pht_num_coherent_state_solution}
    \frac{\partial}{\partial t}\alpha^*_{j_1j_2}(t) = 2\epsilon^*_\td(t)\text{sin}(\omega^*_\td t)e^{i\omega^*_\td t} \\
    -i\Big(\omega_{\R,j_1j_2}(|\alpha^*_{j_1j_2}(t)|^2)- \omega_\td^*\Big)\alpha^*_{j_1j_2}(t)-\frac{\kappa^*}{2}\alpha^*_{j_1j_2}(t),
\end{multline}
where now $\kappa^*=2|\chi_{12}(n_{\tss}^{(2)})|$, $\omega_\td^*=\tilde{\omega}_\R(n^{(2)}_{\tss})$ and $\epsilon_\td^*(t)~=~\sqrt{2n_{\tss}^{(2)} |\chi_{12}(n_{\tss}^{(2)})|^2}$ \footnote{Note that \eqref{eqn:pht_num_coherent_state_solution} defines the coherent state dynamics for all computational transmon states so, in the steady state, $|\alpha^*_{j_1j_2}|^2 =  n_{\tss}^{(2)}$ only if $j_1=j_2$.}. For $j_1\neq j_2$ in the steady state, the effective resonator frequency is closer to the drive frequency and therefore  $|\alpha^*_{j_1j_2}|^2>n_{\tss}^{(2)}$. A consequence of this is that if the CNOT fails and leaves the transmons in $\ket{j_1j_2}$ with $j_1\neq j_2$, then ionization is much more likely. 

In Appendix D we explain how $\omega_{\R,j_1j_2}(n)$ can be numerically determined using the recently introduced method of \emph{branch analysis}, a powerful technique, validated against experimental observations \cite{Shillito_2022,Dumas_2024}. Using $\omega_{\R,j_1j_2}(n)$, \eqref{eqn:pht_num_coherent_state_solution} can be numerically integrated. 

Realistic CNOT gate noise is modelled using a depolarising channel $\Lambda_p$ with depolarising parameter $p$:  $\Lambda_p(\ketbra{j_10}{j_10})=(1-p)\ketbra{j_1j_1}{j_1j_1} + \frac{p}{4}\mathbb{I}_4$. Implementing a noisy CNOT and performing homodyne measurement on the resulting mixed state outputs an integrated voltage signal $V_{\ket{j_1j_1},\tau,p}$ which is a random variable with probability distribution
\begin{multline}\label{eqn:cnot_noise_dist}
\!\!\!\!\!\!\mathcal{P}^*_{\ket{j_1j_1},\tau,p}(v)=\left(1-\frac{3p}{4}\right)\mathcal{P}^*_{\ket{j_1j_1},\tau}(v) + \frac{p}{4}\sideset{}{'}\sum_{j'_1j'_2}\mathcal{P}^*_{\ket{j'_1j'_2},\tau}(v)
\end{multline}
where $\mathcal{P}^*_{\ket{j_1j_2},\tau}(v)$ is the distribution for the integrated voltage of a homodyne measurement of $\ket{\alpha^*_{j_1j_2}(t)}$ (with $\alpha^*_{j_1j_2}(t)$ solutions to \eqref{eqn:pht_num_coherent_state_solution}) and the summation is over $j'_1j'_2 \neq j_1j_1$. Taking into account the photon number dependence and CNOT noise, the SNR \eqref{eqn:snr} becomes 
\begin{equation}\label{eqn:snr_noise}
    S^{(2)}_p(\tau) = S\left(\mathcal{P}^*_{\ket{00},\tau,p}, \mathcal{P}^*_{\ket{11},\tau,p}\right).
\end{equation} 
Further details are given in Appendix A.

The speed-up \eqref{eqn:space-time-trade-off} is derived under the condition that the resonator in both schemes is driven to the same steady-state photon number,  $n_\text{ss}^{(2)}=n_\text{ss}^{(1)}$. Eq.~\eqref{eqn:pht_num_coherent_state_solution} is valid for ``low'' photon numbers, meaning that $n_\text{ss}^{(q)}$ is sufficiently below $n_\crit^{(q)}$ (the precise relation is given in Appendix B). $n_\crit^{(1)}$ and $n_\crit^{(2)}$ will in general not be equal, and if $n_\crit^{(2)}\ll n_\crit^{(1)}$ then the condition $n_\text{ss}^{(2)}=n_\text{ss}^{(1)}$ may not be valid and the speed-up \eqref{eqn:space-time-trade-off} not achievable. We employ branch analysis \cite{Shillito_2022,Dumas_2024}, originally developed for the TR system, and extend it to TTR systems, to estimate  $n_\crit^{(2)}$ numerically and compare it to $n_\crit^{(1)}$. Details are provided in Appendix C. We show that for most of the parameter space analysed $n_\crit^{(2)}$, $n_\crit^{(1)}\geq 35$ is a reasonable lower bound for the critical photon number, from which we can take $n_\text{ss}^{(2)}=n_\text{ss}^{(1)} = 18$ (all choices fully justified in Appendix B).

In Fig.~\ref{fig:snr_s_ratios} we present our main results. To compare our scheme with the standard single transmon dispersive readout we define the ratio $\frac{\tau^{(2)}_6 + \tau_\text{C}}{\tau^{(1)}_6}$ where  $\tau_\text{C}$ is the duration of a CNOT gate, $\tau^{(2)}_6$ is the measurement duration achieving $S^{(2)}_p(\tau^{(2)}_6)=6$ (and analogously for $\tau^{(1)}_6$). 
A speed-up is demonstrated for most parameters explored, allowing for measurement durations as small as a factor of approximately $0.6$. It is most dramatic when $\Delta_1\approx\Delta_2$ but, even when these differ, a significant improvement over the standard readout can still be achieved. 

\subsection{Conclusion}
In this Letter we have presented a scheme for ancilla-assisted speed-up of dispersive readout of a transmon qubit. Under realistic conditions, and over a wide range of parameters, we have demonstrated a significant speed-up can be achieved. We have carefully ensured the applicability of our scheme, employing branch analysis and extending it (itself of independent interest), and incorporated photon number dependence and realistic CNOT noise into our analysis.

Our analysis could be further extended by considering optimising the scheme, e.g.  $n_\crit^{(2)}$ and $\chi_{12}(n)$, over the TR detunings $\Delta_1$ and $\Delta_2$. Indeed, in agreement with \cite{Dumas_2024}, we have found that $n_\crit^{(2)}$ varies considerably with $\Delta_1$ and $\Delta_2$, leaving the possibility of further improvements in readout speed. The TR detunings explored here are also relatively small and close to the edge of what is within the dispersive regime. While we demonstrate faster measurements are possible in this regime, we also expect the speed-up to be even greater for larger detunings.

The faster readout speed is primarily due to the larger resonator decay rate which is a consequence of the larger effective dispersive shift. Not only does this result in a faster readout speed but also in a faster post measurement ringdown time of the resonator, which further improves the rate at which measurements can be performed.

Our model could be made more realistic by including the tunable coupler and transmon relaxation. We also note that the measurement fidelity \cite{Blais_2021}, an alternative figure of merit to the SNR, will still witness the speed advantage of our scheme but will exhibit a noise floor on its maximum achievable value due to the imperfect CNOT gate.

While our scheme can be seen as an extension of the standard dispersive readout method, our proposal of coupling an additional ancillary transmon to the resonator should also apply to new circuit designs for improved readout via non-perturbative coupling \cite{Chapple_2025, Dassonneville_2020, Beaulieu_2026, Mori_2025, Chapple_2025_a, Ye_2024, Salunkhe_2025}. For example, the two transmons in our scheme could be coupled to the resonator using the balanced cross-Kerr coupling, taking advantage of both the non-perturbative interaction and the space-time trade-off from our scheme.

\subsection{Acknowledgments} The authors are grateful to Carys Harvey, Aleksandr Grigorev, Oliver Green, Christopher Woodgate, Stephen Clark, Gerard Milburn and Alexandre Blaise for insightful discussions.
This work was carried out using the computational facilities of the Advanced Computing Research Centre, University of Bristol - http://www.bristol.ac.uk/acrc/,
and the Isambard 3 Tier-2 HPC Facility. Isambard 3 is hosted by the University of Bristol and operated by the GW4 Alliance (https://gw4.ac.uk) and is funded by UK Research and Innovation; and the Engineering and Physical Sciences Research Council [EP/X039137/1].
We finally acknowledge support from the EPSRC through the grant EP/L015730/1.

\subsection{AI Usage statement} AI tools (Anthropic Claude Opus 5) were used to help code the simulations, as a final check of calculations performed in the manuscript and for general copy-editing. 

\setcounter{equation}{0}
\renewcommand{\theequation}{A\arabic{equation}}

\subsection{Appendix A: SNR for a noisy CNOT gate}
The mean and variance under the depolarising CNOT noise entering \eqref{eqn:snr_noise} are given by 
\begin{multline}
    \!\!\!\!\expV{V}{\mathcal{P}^*_{\ket{j_1j_2},\tau,p}} = \left(1-\frac{3p}{4}\right)\expV{V}{\mathcal{P}^*_{\ket{j_1j_2},\tau}}  + \frac{p}{4}\sideset{}{'}\sum_{j'_1j'_2}\expV{V}{\mathcal{P}^*_{\ket{j'_1j'_2},\tau}}
\end{multline}
and
\begin{multline}
    \varV{V}{\mathcal{P}^*_{\ket{j_1j_2},\tau,p}} = \left(1-\frac{3p}{4}\right)\left(\varV{V}{\mathcal{P}^*_{\ket{j_1j_2},\tau}}+\expV{V}{\mathcal{P}^*_{\ket{j_1j_2},\tau}}^2\right) \\
    +\frac{p}{4}\sideset{}{'}\sum_{j'_1j'_2}\left(\varV{V}{\mathcal{P}^*_{\ket{j'_1j'_2},\tau}}+\expV{V}{\mathcal{P}^*_{\ket{j'_1j'_2},\tau}}^2\right) - \expV{V}{\mathcal{P}^*_{\ket{j_1j_2},\tau,p}}^2.
\end{multline}
where the summation is over $j'_1j'_2\neq j_1j_2$. Expressions for the means and variances are given in SM D. 

\setcounter{equation}{0}
\renewcommand{\theequation}{B\arabic{equation}}

\subsection{Appendix B: Steady state photon number and SNR threshold}
$n_\text{ss}^{(q)}$ is as defined in \cite{Chapple_2025}. As discussed in the main text the resonator is approximately a coherent state so its photon number distribution follows a Poisson distribution. In the steady state, the mean and variance of the photon number distribution are therefore $n_\text{ss}^{(q)}$. We define $n_\text{ss}^{(q)}$ such that the coherent state tail has a minimal overlap with the critical photon number. An overlap of $0.01\%$ is considered sufficient which corresponds to 4 standard deviations above the mean. Therefore $n_\text{ss}^{(q)}=\left(\sqrt{n_\crit^{(q)}+4}-2\right)^2$.

Following \cite{Chapple_2025}, we choose $n_\crit^{(q)}=35$, which corresponds to $n_\text{ss}^{(q)}=18$. They justify this as a readout duration of $\tau\sim\kappa^{-1}$, which approximately takes the resonator to its steady state, corresponds to an SNR of $S^{(q)}(\tau)\sim6$ (if $\eta=0.5$, in line with state of the art experiments) which is considered ``good''.
\begin{figure}[t!]
    \centering
    \includegraphics[width=\linewidth]{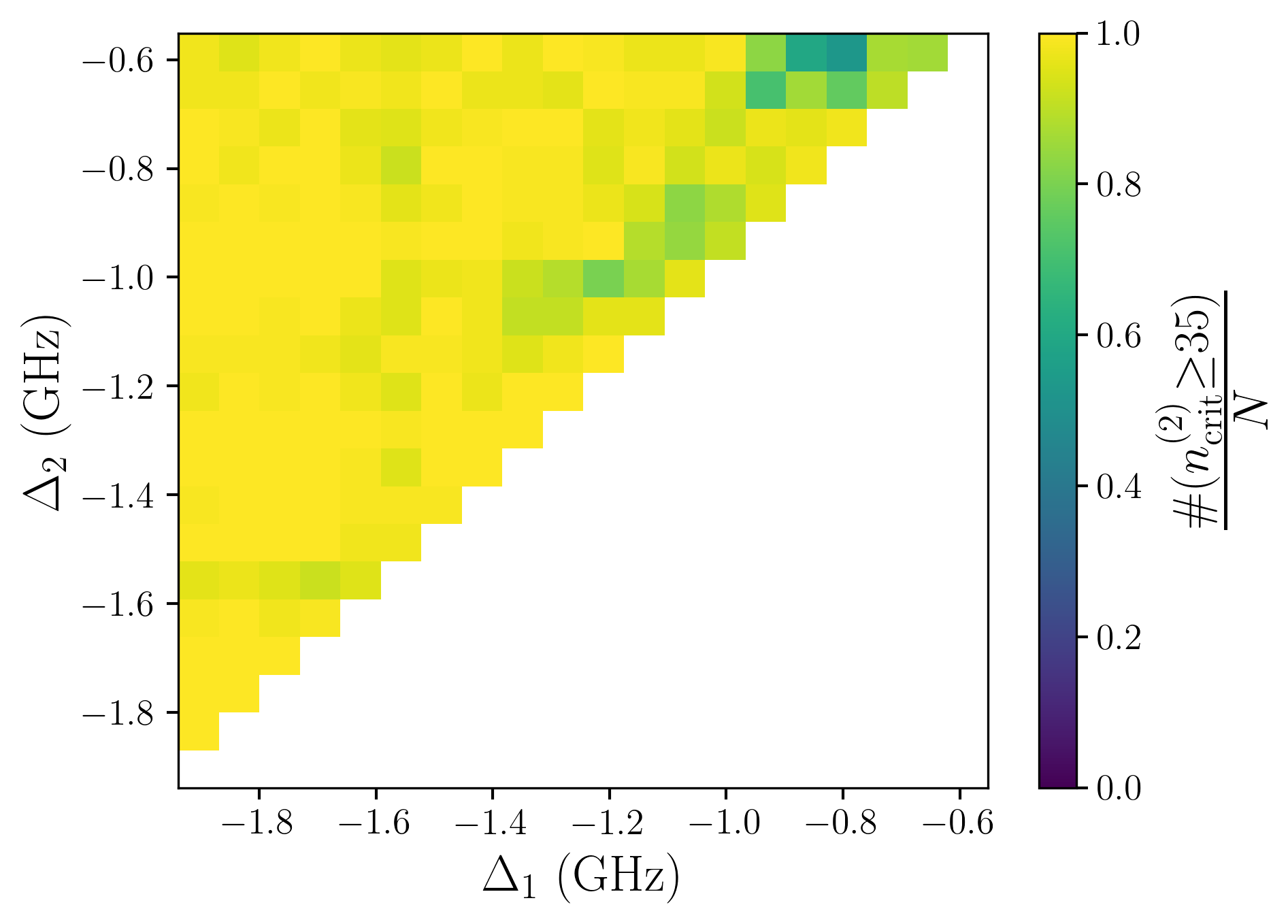}
    \caption{\textbf{TTR system branch analysis results.} This figure uses the same system parameters as Fig. \ref{fig:snr_s_ratios} except for $n_{\g,1}$ and $n_{\g,2}$. Each data point is the fraction of samples for which $n_\crit^{(2)}\geq35$ where $n_\crit^{(2)}$ was sampled $N=100$ times over uniform samples of the pair $n_{\g,1},n_{\g,2}\in[0,0.5]$ using semi-classical branch analysis on the TTR system. The same analysis was performed for the equivalent TR system by taking $N$ samples of $n_{\g}\in[0,0.5]$ for each data point. It was found that the fraction of samples for which $n_\crit^{(1)}\geq35$ was almost always $1$ with the lowest fraction being $0.97$. This means $n_\crit^{(1)}=35$ is a good lower bound for the TR system. For the TTR system, most of the parameter space simulated suggests $n_\crit^{(2)}=35$ is a good lower bound. Points in the top right corner, towards $\Delta_1\approx\Delta_2\approx-0.6\,\text{GHz}$, tend to have lower percentages suggesting this will not be a good regime for our scheme.}
    \label{fig:two_tmon_n_crit_violations}
\end{figure}

\setcounter{equation}{0}
\renewcommand{\theequation}{C\arabic{equation}}

\subsection{Appendix C: Branch analysis}
As succinctly summarised in \cite{Chapple_2025}, for a single transmon system, when the limits of the dispersive regime are pushed, readout is compromised by multiphoton processes that emerge from degeneracies between the transmon and resonator energy levels. This phenomenon is known as ionization and arises when two eigenstates of the TR system become resonant as photons populate the resonator, resulting in an abrupt transfer of population from the resonator to the transmon. The resonator’s critical photon number $n_\crit^{(1)}$ is the number of photons above which these multiphoton processes first start to occur and it is highly dependent on the system parameters. Critical photon numbers can be evaluated using the recently introduced branch analysis; a powerful numerical technique, validated against experimental observations, that accurately predicts the onset of ionization \cite{Dumas_2024, Shillito_2022}.

To perform branch analysis begin with the undriven transmon and resonator Hamiltonians, $H_\T$ and $H_\R$ respectively, and the coupling between them $V$, generically given by $H_\TR = H_\T + H_\R + V$. Diagonalising this Hamiltonian results in a set of eigenvectors $\{\ket{\lambda}\}$. For each transmon level $j$, find the eigenvector from $\{\ket{\lambda}\}$ that maximizes the overlap $|\braket{j,0}{\lambda}|^2$ (with the resonator in state $\ket{0}$) and label this eigenvector as $\overline{\ket{j,0}}$, the overline denoting a dressed state. Then, using these states $\overline{\ket{j,0}}$, recursively assign a label to all remaining states where $\overline{\ket{j,n+1}}$ is defined as the unassigned eigenstate $\{\ket{\lambda}\}$ which maximizes the overlap $|\bra{\lambda}a^\dagger\overline{\ket{j,n}}|^2$. Repeating this for each transmon index $j$ results in a set of dressed states $B_{j} = \{\overline{\ket{j,n}}\}_{n\in\mathbf{N}}$ which are referred to as branches.
Using these branches, the average transmon and resonator populations can be computed. An ionization process can be identified when the branches for the ground or first excited state swap with a branch of a higher excited state. The photon number $n$ at which this occurs is reported as the critical photon number.

Branch analysis was only developed for the TR system and $n_\crit^{(1)}$ cannot simply be applied to the TTR system since the transmons are coupled via the resonator. This means resonances between different transmon states can be mediated by the resonator, resulting in an abrupt transfer of population between the two transmons, a physical effect not present in the TR system. 

Branch analysis, in theory, also extends to the TTR system. However, the size of the computation quickly becomes intractable due to the size of the Hilbert space that needs to be processed. 

A more computationally efficient semi-classical branch analysis method was developed in \cite{Dumas_2024} for the TR system which traces out the resonator subsystem while maintaining the effect of the driven resonator on the transmon. This semi-classical method gives consistent results for $n_\crit^{(1)}$ when compared against the fully quantum method which includes both the transmon and the resonator.

We extend the semi-classical model to the TTR system, however, we must be careful when tracing over the resonator subsystem in the derivation of the semi-classical model. As previously mentioned, the transmons are coupled via the resonator so an effective coupling must be found before the resonator can be traced out. This effective coupling is a novel addition to the previously developed semi-classical model. For further details on branch analysis for the TTR system see SM E.

Similarly to the TR system, for a given set of parameters the critical photon number varies chaotically with the transmon gate charges $\{n_{\g,1}, n_{\g,2}\}$ so $n_\crit^{(2)}$ cannot simply be calculated for a set of parameters. Instead, for a given set of parameters, $n_\crit^{(2)}$ is calculated many times over random samples of $\{n_{\g,1},n_{\g,2}\}$ and the percentage of outcomes which are above a desired lower bound are quoted, as in Fig. \ref{fig:two_tmon_n_crit_violations} where we lower bound $n_\crit^{(2)}$ at 35.

We should bear in mind that the estimates of $n_\crit^{(2)}$ are heuristic and are not currently backed up by experimental evidence. However, they have been calculated using an extension of branch analysis, currently the most cutting-edge tool for understanding the limits of the dispersive regime, barring experimental results. 

\setcounter{equation}{0}
\renewcommand{\theequation}{D\arabic{equation}}

\subsection{Appendix D: Photon Number Dependent Resonator Frequency}
The resonator frequency is often defined as $\omega_{\R,j_1j_2}=E_{j_1j_2,1}-E_{j_1j_2,0}$ when the transmons are in state $\ket{j_1j_2}$ with $E_{j_1j_2,n}$ the energy of the dressed state $\overline{\ket{j_1j_2,n}}$ as described in Appendix C. However, due to the coupling with the transmons, the resonator is no longer a harmonic oscillator so the energy gap changes with photon number $E_{j_1j_2,1}-E_{j_1j_2,0}\neq E_{j_1j_2,n+1}-E_{j_1j_2,n}$ \cite{Chapple_2025, Gambetta_2006, Schuster_2005}. We therefore define the photon number dependent resonator frequency to be
\begin{equation}
    \omega_{\R,j_1j_2}(n)=E_{j_1j_2,n+1}-E_{j_1j_2,n}
\end{equation}
which is the definition used in \eqref{eqn:pht_num_coherent_state_solution}, where $E_{j_1j_2,n}$ is calculated using the semi-classical branch analysis method described in Appendix C.

\bibliography{main}

\setcounter{equation}{0}
\renewcommand{\theequation}{S\arabic{equation}}
\renewcommand{\thefigure}{S\arabic{figure}}

\newpage\onecolumngrid
\section{Supplemental Material}

\subsection{Note on References}
Figure and equation references prefixed with S refer to the Supplemental Material (SM). If a reference has another prefix or no prefix then it refers to the main text.

\subsection{A - Review of the Dispersive Hamiltonian for a Transmon-Resonator System}
In this section we review the derivation of the dispersive Hamiltonian for a single transmon-resonator (TR) system \eqref{eqn:H_disp_s} to set the stage for SM B where we perform the derivation of the dispersive Hamiltonian for a transmon-transmon-resonator (TTR) system. 

The full Hamiltonian for the TR system is ($\hbar=1$) \cite{Koch_2007, Blais_2021, Krantz_2019}
\begin{align}
    H_{\TR} &= H_\T + H_\R + V \label{eqn:htr_1_S}\\
    H_{\T} &= 4E_{\C}\left(n_{\T}-n_{\g}\right)^2 - E_{\J}\text{cos}\left(\phi_{\T}\right)  \label{eqn:bare_ham_tmon} \\
    H_{\R} &= \omega_{\R}a^\dagger a \label{eqn:bare_ham_res} \\
    V &= - ig \left(n_{\T}-n_{\g}\right) (a-a^\dagger) \label{eqn:coupling_term}
\end{align}
where $E_{\C}$ is the charging energy, $E_{\J}$ the Josephson energy, $n_{\g}$ the gate charge constant, $\omega_{\R}$ the resonator frequency, $g$ the coupling strength, $n_{\T}$ and $\phi_{\T}$ the transmon charge and phase conjugate operators, and $a$ the resonator annihilation operator. 

The system is operated in the transmon regime $E_{\J}/E_{\C}\gg1$ which suppresses the effect of the gate charge $n_{\g}$ in the qubit subspace and, as such, it is standard to drop the $n_g$ term in \eqref{eqn:coupling_term} to simplify the analysis \cite{Koch_2007}. We therefore replace $V$ in \eqref{eqn:htr_1_S} with $V'=-ig n_{\T}(a-a^\dagger)$.

To understand the physics in the low energy regime we make the Schrieffer-Wolff transformation (SWT) \cite{Bravyi_2011, Blais_2004}. To do this, we split \eqref{eqn:htr_1_S} into the bare uncoupled Hamiltonian term $H_0$ and the coupling term $V'$ such that $H'_{\TR}=H_0+V'$ with
\begin{align}\label{eqn:ham-pert}
    H_0 &= H_\T + H_\R = \sum_{j} E_j\ket{j}\bra{j} + \omega_\R a^\dag a, \nonumber \\
    V' &= - i\sum_{i,j} g_{ij}\ket{i}\bra{j}(a - a^\dag) 
\end{align}
where we have expressed the transmon operators in terms of the eigenstates $\ket{j}$ of $H_\T$ such that $H_\T\ket{j}=E_j\ket{j}$ and $g_{ij}=g\bra{i}n_\T\ket{j}$.

Consider the unitary transformation $e^{S}$ where $S$ is an anti-Hermitian operator ${S^\dag=-S}$ yet to be determined. We use this unitary operator to transform the Hamiltonian $H'_{\TR}$ and expand the result 
\begin{equation}\label{eqn:SW-expansion}
    \tilde{H}_{\TR}=e^{S}H'_{\TR}e^{-S}=H_0+V'+[S,H_0]+[S,V']+\frac{1}{2}[S,[S,H_0]]+\mathcal{O}(g^3).
\end{equation}
By choosing $S$ such that $[H_0,S]=V'$ the second and third terms cancel and the fifth term becomes the same order as the fourth term. So, to second order in $g$, the SWT is 
\begin{equation}\label{eqn:SWT}
    \tilde{H}_{\TR}=H_0+\frac{1}{2}[S,V']+\mathcal{O}(g^3).
\end{equation}

We define $S=-i(S_\T a^\dag+S^\dag_\T a)$ where $S_\T=\sum_{i,j}\frac{g_{ij}}{\omega_{ij}-\omega_\R}\ket{i}\bra{j}$ with $\omega_{ij}=E_j-E_i$ and, using $H_0$ and $V'$ from \eqref{eqn:ham-pert}, we can show that $[H_0,S]=V'$. Substituting into \eqref{eqn:SWT} we find
\begin{align}\label{eqn:H_eff}
    \tilde{H}_{\TR} &= \sum_{j} \left(E_j + \Lambda_{jj}\right)\ket{j}\bra{j} + \left(\omega_\R + \sum_{j}\chi_{jj}\ket{j}\bra{j}\right)a^\dag a \nonumber \\
    &\quad + \sum_{\substack{i,j \\ i\neq j}} \Lambda_{ij}\ket{i}\bra{j} + \sum_{\substack{i,j \\ i\neq j}} \chi_{ij}\ket{i}\bra{j}a^\dag a \nonumber \\
    &\quad + \left(\sum_{i,j} \eta_{ij}\ket{i}\bra{j}a^\dag a^\dag + \text{h.c.}\right) + \mathcal{O}(g^3)
\end{align}
where
\begin{align}
    \Lambda_{ij} &= -\frac{1}{2}\sum_{k} g_{ik}g_{kj}\left(\frac{1}{\omega_\R-\omega_{kj}}+\frac{1}{\omega_\R-\omega_{ki}}\right) \label{eqn:lamb_shift}\\
    \chi_{ij} &= \sum_{k} g_{ik}g_{kj}\left(\frac{\omega_{ik}}{\omega_\R^2-\omega_{ik}^2} + \frac{\omega_{jk}}{\omega_\R^2-\omega_{jk}^2}\right) \label{eqn:disp_shift}\\
    \eta_{ij} &= \frac{1}{2}\sum_{k} g_{ik}g_{kj}\left(\frac{1}{\omega_\R-\omega_{kj}} - \frac{1}{\omega_\R+\omega_{ki}}\right). \label{eqn:2photon_shift}
\end{align}

In the dispersive regime $\frac{g}{|\omega_{01}-\omega_\R|}\ll1$ is a small quantity so to make the dispersive approximation we drop $\mathcal{O}(g^3)$ from the above Hamiltonian as it is considered to have minimal effect on the dynamics. It is also standard to only keep the first line of \eqref{eqn:H_eff} and drop the off-diagonal terms in the last two lines as they are much smaller than the energy gaps between the lower transmon energy levels so have a negligible effect on the dynamics within the computational subspace. This is motivated by applying the rotating wave approximation \cite{Burgarth_2022, Burgarth_2024}.
 
We simplify the Hamiltonian further by truncating to the qubit subspace; the first two energy levels of the transmon. This results in the dispersive Hamiltonian for qubit measurement
\begin{equation}\label{eqn:H_disp_s}
    \tilde{H}_{\disp}=\frac{\omega_{\T}}{2}\sigma_{Z} + \left(\tilde{\omega}_{\R} + \chi\sigma_{Z}\right)a^\dagger a
\end{equation}
where $\omega_{\T}=\omega_{01}+\Lambda_{11}-\Lambda_{00}$ is the transmon's qubit subspace frequency, $\tilde{\omega}_{\R}=\omega_\R + \frac{\chi_{00}+\chi_{11}}{2}$, $\chi=\frac{\chi_{11}-\chi_{00}}{2}$, $\sigma_Z=\ket{1}\bra{1}-\ket{0}\bra{0}$ and we have ignored terms which are proportional to identity. The first term shows how the qubit frequency $\omega_{01}$ is shifted by the coupling to the resonator, while the second term shows how the resonator frequency is dependent upon the qubit state. 

\subsection{B - Derivation of the Dispersive Master Equation for a Transmon-Transmon-Resonator System}
In this section we derive the dispersive Hamiltonian for a TTR system and then introduce a resonator drive and decay term in order to construct a Master equation for modelling dispersive readout as expressed in \eqref{eqn:lindblad_disp_S}. 

The equation for the TTR system is described by \eqref{eqn:htr} in the main text, reproduced here for reference, 
\begin{align}
    H_{\TTR} &= \sum_{q=1}^2 H_{\T,q} + H_\R + \sum_{q=1}^2V_q \label{eqn:htr_S}\\
    H_{\T,q} &= 4E_{\C,q}\left(n_{\T,q}-n_{\g,q}\right)^2 - E_{\J,q}\text{cos}\left(\phi_{\T,q}\right) \label{eqn:htr_S_1}\\
    H_\R &= \omega_{\R}a^\dagger a \label{eqn:htr_S_2}\\
    V_{q} &= - ig_{q} \left(n_{\T,q}-n_{\g,q}\right) (a-a^\dagger). \label{eqn:coupling_term_2}
\end{align}
This is the direct analogue of \eqref{eqn:htr_1_S} but now with two transmons independently coupled to the resonator. The parameters are as described for \eqref{eqn:htr_1_S} but with the subscript $q\in\{1,2\}$ specifying to which transmon a parameter belongs. 

As in SM A we drop $n_{\g,q}$ from \eqref{eqn:coupling_term_2}, replacing $V_q$ with $V'_q=-ig_{q} n_{\T,q} (a-a^\dagger)$, and we split the Hamiltonian into its bare $H_0$ and coupling $V_q$ terms such that
\begin{align}
    H'_{\TTR}& =H_0+\sum_{q=1}^2V'_q \label{eqn:simplified_coupling_TTR} \\
    H_0  &= \sum_{q=1}^2H_{\T,q} + H_\R = \sum_{q=1}^2\sum_{j} E_{j,q}\ket{j_q}\bra{j_q} +\omega_\R a^\dag a, \\
    V'_q &= -i\sum_{i,j} g_{ij,q}\ket{i_q}\bra{j_q}(a-a^\dag). \label{eqn:2_tmon_couple}
\end{align}
where we have expressed the transmon operators in terms of the eigenstates $\ket{j_q}$ of $H_{\T,q}$ such that $H_{\T,q}\ket{j_q}=E_{j,q}\ket{j_q}$ and $g_{ij,q}=g_q\bra{i_q}n_{\T,q}\ket{j_q}$.

We perform the SWT on $H'_{\TTR}$ to find
\begin{equation}\label{eqn:SW-expansion_2}
    \tilde{H}_{\TTR}=e^{S}H'_{\TTR}e^{-S}=H_0+\sum_{q=1}^2V'_q+[S,H_0]+\sum_{q=1}^2[S,V'_q]+\frac{1}{2}[S,[S,H_0]]+\mathcal{O}(g^3).
\end{equation}
If we choose 
\begin{equation}\label{eqn:SWT_operator}
    S=\sum^2_{q=1}S_q    
\end{equation}
with $S_q=-i(S_{\T,q}a^\dag+S_{\T,q}^\dag a)$ and $S_{\T,q}=\sum_{i,j}\frac{g_{ij,q}}{\omega_{ij,q}-\omega_\R}\ket{i_q}\bra{j_q}$ we find $[H_0,S_q]=V'_q$ and therefore
\begin{align}\label{eqn:SWT-expansion-2}
    \tilde{H}_{\TTR} &=  H_0 + \frac{1}{2}\sum_{q,p=1}^2[S_q,V'_p] + \mathcal{O}\left(g^3\right), \nonumber\\
    &=\omega_\R a^\dag a + \sum_{q=1}^2\left(\sum_{j} \left(E_{j,q} + \Lambda_{jj,q}\right)\ket{j_q}\bra{j_q} + \sum_{j}\chi_{jj,q}\ket{j_q}\bra{j_q}a^\dag a\right. \nonumber \\
    &\quad + \sum_{\substack{i,j \\ i\neq j}} \Lambda_{ij,q}\ket{i_q}\bra{j_q} + \sum_{\substack{i,j \\ i\neq j}} \chi_{ij,q}\ket{i_q}\bra{j_q}a^\dag a \nonumber \\
    &\quad \left.+ \sum_{i,j} \left(\eta_{ij,q}\ket{i_q}\bra{j_q}a^\dag a^\dag + \text{h.c.}\right)\right) \nonumber \\
    &\quad + \frac{1}{2}\sum_{\substack{i,j\\k,l}} J_{ijkl}\ket{i_1k_2}\bra{j_1l_2} + \mathcal{O}(g^3)
\end{align}
where $\Lambda_{ij,q}$, $\chi_{jj,q}$, $\eta_{ij,q}$ are the parameters defined in \eqref{eqn:lamb_shift}, \eqref{eqn:disp_shift} and \eqref{eqn:2photon_shift} respectively for transmon $q$ and 
\begin{equation}
    J_{ijkl}=g_{ij,1}g_{kl,2}\left(\frac{1}{\omega_{ij,1}-\omega_\R} + \frac{1}{\omega_{ji,1}-\omega_\R} + \frac{1}{\omega_{kl,2}-\omega_\R} + \frac{1}{\omega_{lk,2}-\omega_\R}\right)    
\end{equation}
is the coupling strength between transmons 1 and 2 induced by their respective coupling to the resonator. We make the dispersive approximation by dropping $\mathcal{O}(g^3)$ and neglect the single transmon off-diagonal terms. The final term in \eqref{eqn:SWT-expansion-2} includes exchange terms between the different transmon states. 

After dropping the single transmon off-diagonal terms and then truncating the transmons to their qubit subspace, we apply the RWA to the transmon-transmon coupling, allowing us to neglect the fast rotating terms because they have negligible effect on the dynamics \cite{Burgarth_2022, Burgarth_2024}. After applying the RWA, the remaining terms couple the states $\ket{01}$ and $\ket{10}$. We now find the approximate Hamiltonian for the qubit subspace
\begin{equation}
    \tilde{H}_{\disp}=\sum_{q=1,2}\frac{\omega_{\T,q}}{2}\sigma_{Z,q} + \left(\tilde{\omega}_{\R} + \sum_{q=1,2}\chi_q\sigma_{Z,q}\right)a^\dagger a  +  J\sigma_{+,1}\sigma_{-,2} + J^*\sigma_{-,1}\sigma_{+,2} - \frac{2}{\omega_\R}G_1G_2
\end{equation}
where $\omega_{\T,q}=\omega_{01,q}+\Lambda_{11,q}-\Lambda_{00,q}$, $\tilde{\omega}_{\R}=\omega_\R + \sum_{q=1}^2\frac{\chi_{00,q}+\chi_{11,q}}{2}$, $\chi_q=\frac{\chi_{11,q}-\chi_{00,q}}{2}$, $J=\frac{g_{10,1}g_{01,2}}{2}\sum_{q=1}^2\left(\frac{1}{\omega_{10,q}-\omega_\R}+\frac{1}{\omega_{01,q}-\omega_\R}\right)$, $\sigma_{+,q}$ and $\sigma_{-,q}$ are the raising and lowering operators and $G_q=g_{00,q}\ketbra{0_q}{0_q} + g_{11,q}\ketbra{1_q}{1_q}$. It can be seen numerically that $g_{00,q}\approx g_{11,q}$ within the transmon regime \cite{Groszkowski_2021} so the final term is approximately identity times a scalar. It therefore has negligible effect on the dynamics and can be dropped.

If we neglect the final transmon-transmon coupling term (the justification of which we discuss below) we recover \eqref{eqn:H_disp} from the main text 
\begin{equation}\label{eqn:H_disp_2_S}
    \tilde{H}_{\disp}=\sum_{q=1,2}\frac{\omega_{\T,q}}{2}\sigma_{Z,q} + \left(\tilde{\omega}_{\R} + \sum_{q=1,2}\chi_q\sigma_{Z,q}\right)a^\dagger a, 
\end{equation}
where we note that, for simplicity of notation in the main text, we have omitted the tilde on $\tilde{H}_\disp$, and denoted the Hamiltonian simply as $H_\disp$. We will continue to use this convention in all that follows. 

We now justify neglecting the coupling term. As discussed in the main text, we assume that a tunable coupler is used to couple the transmons which allows us to implement a CNOT gate without needing to model a direct coupling between them. Additionally, the coupler can be precisely tuned to suppress the indirect coupling between the transmons, facilitated by the resonator. This is a standard technique in superconducting qubit experiments, e.g. \cite{McKay_2016, Yan_2018}. Thus in our current analysis we neglect the coupling term, but we note that in a more complete analysis of our scheme, the effect of the tunable coupler and the indirect coupling term should be included.

To perform a measurement we must drive the resonator. This requires introducing the time dependent drive term \eqref{eqn:drive}, reproduced here for reference,
\begin{equation}\label{eqn:drive_S}
    H_{\td}(t) = -i2\epsilon_\td(t)\text{sin}(\omega_\td t)(a - a^\dag)
\end{equation}
where $\epsilon_\td(t)$ is the drive amplitude and $\omega_\td$ the drive frequency. 

Under the SWT \eqref{eqn:SW-expansion_2}, to first order in $g$, the change to the drive term is
\begin{align}\label{eqn:qubit_drive}
    \sum_{q=1}^2 [S_q,H_\td(t)] 
    = 4\epsilon_\td(t)\sin(\omega_\td t)\sum_{q=1}^2\left(\frac{\omega_\R}{(\omega_{10,q}-\omega_\R)(\omega_{10,q}+\omega_\R)}\left(g_{10,q}\sigma_{+,q} + g_{01,q}\sigma_{-,q}\right) - \frac{1}{\omega_\R}G_q\right)
\end{align}
where we have truncated to the qubit subspace. This term demonstrates that we can drive the qubits via the resonator, however, the drive is first order in $g$ so it must be strong and on resonance with a qubit frequency in order to be relevant. We drive at the resonator frequency, not the transmon frequencies, so in what follows we do not consider the effect of the resonator drive on the transmons. This means the change \eqref{eqn:qubit_drive} is not considered in the final expression for the dispersive Master equation \eqref{eqn:lindblad_disp_S}.

Combining \eqref{eqn:H_disp_2_S} and \eqref{eqn:drive_S} and considering resonator decay at a rate $\kappa$ we obtain the Master equation \eqref{eqn:lindblad_disp}, from the main text, in the Schrieffer-Wolff frame
\begin{equation}\label{eqn:lindblad_disp_S}
    \frac{\partial}{\partial t}\rho = -i[\tilde{H}_{\disp} + H_{\td}(t),\rho] + \kappa\mathcal{D}[a]\rho
\end{equation}
where $\mathcal{D}[a]\rho = a\rho a^\dagger - \{a^\dagger a, \rho\}/2$ is the Lindblad dissipator. 

Finally we consider the effect of the SWT on the resonator decay. The transformation results in an additional decay term on the transmons, known as Purcell decay, which is second order in $g$. Purcell decay can be understood by considering the dressed eigenstates of the TTR system $\overline{\ket{j_1j_2,n}}\approx e^{-S}\ket{j_1j_2,n}$ which results in a small amount of entanglement between the transmons and the resonator, therefore any external coupling to the resonator will also affect the transmons. With careful treatment of the resonator multi-mode transmission line, and by modelling the resonator as a multi-mode cavity, it can be shown that the decay of the transmon via the resonator is at the transmon frequency instead of the resonator frequency \cite{Boissonneault_2009}. By including a Purcell filter which filters both transmon frequencies, the transmon decay can be suppressed without suppressing the resonator decay which should not be suppressed as it needs to be large in order to facilitate measurement \cite{Blais_2021, Krantz_2019, Sete_2015}.

\subsection{C - Solution for the Dispersive Master Equation}
In this section we use \eqref{eqn:lindblad_disp_S} to find the dynamical equation for the evolution of the resonator state \eqref{eqn:diff_A} as it is driven recovering \eqref{eqn:coherent_state_solution} in the main text. We then solve the dynamical equation under the RWA to find a solution for the coherent state \eqref{eqn:coherent_state} which will be used in the next section to find the SNR. For completeness, we also find a solution for the coherences and amplitudes of the full quantum system \eqref{eqn:amplitude}.

In order to solve \eqref{eqn:lindblad_disp_S} we first transform the Master equation into the frame rotating at the qubit and resonator drive frequency using the time dependent unitary $U(t)=\text{exp}(i(\sum_{q=1}^2\frac{\omega_{\T,q}}{2}\sigma_{Z,q}+\omega_\td a^\dag a)t)$. In this rotating frame the Master equation is
\begin{align}\label{eqn:master_equation_S}
    \frac{\partial}{\partial t}\tilde{\rho} =\mathcal{L}[\tilde{\rho}] = & -i[(\Delta_\td+\sum_{q=1}^2\chi_q\sigma_{Z,q})a^\dag a,\tilde{\rho}] - 2\epsilon_\td(t)\text{sin}(\omega_\td t)[e^{-i\omega_\td t}a - e^{i\omega_\td t}a^\dag, \tilde{\rho}] + \kappa(a\tilde{\rho}a^\dag - \frac{1}{2}\{a^\dag a,\tilde{\rho}\})
\end{align}
where $\tilde{\rho}(t)=U(t)\rho(t)U^\dag(t)$, $\mathcal{L}[\cdot]$ is the Liouville operator and $\Delta_\td=\tilde{\omega}_\R-\omega_\td$.

To solve this equation we follow a similar derivation to \cite{Gambetta_2006}. We start with a trial solution
\begin{equation}\label{eqn:trial_state}
    \tilde{\rho}(t) = \sum_{\mu,\nu\in\{0,1\}^{\otimes 2}}C_{\mu\nu}(t)\ket{\mu}\bra{\nu}\otimes\ket{\alpha_\mu(t)}\bra{\alpha_\nu(t)}
\end{equation}
where $\mu$ ($\nu$) is a bit string of length $2$ and $\ket{\mu}$ ($\ket{\nu}$) is the corresponding state of the $2$ qubits, $C_{\mu\nu}(t)$ is a time dependent amplitude and $\ket{\alpha(t)}=e^{-|\alpha(t)|^2/2}\sum_{n=0}^\infty\frac{\alpha(t)^n}{\sqrt{n!}}\ket{n}$ is a time dependent coherent state. This trial solution is motivated by the drive applied to the resonator, which generates a coherent state in an isolated system, and the resonator decay channel, which transforms a coherent state to another coherent state. Since the Master equation is linear, we will solve it for an arbitrary term $\tilde{\rho}_{\mu\nu}(t)=C_{\mu\nu}(t)\ket{\mu}\bra{\nu}\otimes\ket{\alpha_\mu(t)}\bra{\alpha_\nu(t)}$. For brevity in what follows we do not explicitly write the time dependence of $\ket{\alpha_\mu(t)}$ or $C_{\mu\nu}(t)$. The differential of a coherent state with respect to time is 
\begin{align}
    \frac{\partial}{\partial t}\ket{\alpha} = e^{-\frac{|\alpha|^2}{2}}\sum_{n=0}^\infty\frac{n\alpha^{n-1}\left(\frac{\partial}{\partial t}\alpha\right)}{\sqrt{n!}}\ket{n} - \frac{1}{2}\frac{\partial}{\partial t}\left(|\alpha|^2\right)\ket{\alpha} 
    = \left(\left(\frac{\partial}{\partial t}\alpha\right)a^\dag - \frac{1}{2}\frac{\partial}{\partial t}\left(|\alpha|^2\right)\right)\ket{\alpha}
\end{align}
therefore the LHS of \eqref{eqn:master_equation_S} gives 
\begin{align}\label{eqn:master_LHS}
    \frac{\partial}{\partial t}\rho_{\mu\nu} &= \ket{\mu}\bra{\nu}\otimes\left(C_{\mu\nu}\left(\frac{\partial}{\partial t}\ket{\alpha_\mu}\right)\bra{\alpha_\nu} + C_{\mu\nu}\ket{\alpha_\mu}\left(\frac{\partial}{\partial t}\bra{\alpha_\nu}\right) + \frac{\partial}{\partial t}C_{\mu\nu}\ket{\alpha_\mu}\bra{\alpha_\nu}\right) \nonumber \\
    &= \left(-\frac{1}{2}\frac{\partial}{\partial t}\left(|\alpha_\mu|^2\right)-\frac{1}{2}\frac{\partial}{\partial t}\left(|\alpha_\nu|^2\right)+\frac{\frac{\partial}{\partial t}C_{\mu\nu}}{C_{\mu\nu}}\right)\tilde{\rho}_{\mu\nu} \nonumber \\
    &\quad+ \left(\frac{\partial}{\partial t}\alpha_\mu\right)a^\dag\tilde{\rho}_{\mu\nu} \nonumber \\
    &\quad + \left(\frac{\partial}{\partial t}\alpha_\nu^*\right)\tilde{\rho}_{\mu\nu}a.
\end{align}
Recalling $a\ket{\alpha}=\alpha\ket{\alpha}$, the RHS of \eqref{eqn:master_equation_S} gives
\begin{align}\label{eqn:master_RHS}
    \mathcal{L}[\rho_{\mu\nu}] &= (-2\epsilon_\td(t)\text{sin}(\omega_\td t)(e^{-i\omega_\td t}\alpha_\mu+e^{i\omega_\td t}\alpha_\nu^*)+\kappa\alpha_\mu\alpha_\nu^*)\tilde{\rho}_{\mu\nu} \nonumber \\
    &\quad + (2\epsilon_\td(t)\text{sin}(\omega_\td t)e^{i\omega_\td t}-i(\Delta_\td+\tilde{\chi}_\mu)\alpha_\mu -\frac{\kappa}{2}\alpha_\mu)a^\dag\tilde{\rho}_{\mu\nu} \nonumber \\
    &\quad + (2\epsilon_\td(t)\text{sin}(\omega_\td t)e^{-i\omega_\td t} + i(\Delta_\td+\tilde{\chi}_\nu)\alpha_\nu^* -\frac{\kappa}{2}\alpha_\nu^*)\tilde{\rho}_{\mu\nu}a
\end{align}
where 
\begin{equation}\label{eqn:effective_chi}
    \tilde{\chi}_\mu = \bra{\mu}\sum_{q=1}^2\chi_q\sigma_{Z,q}\ket{\mu}.    
\end{equation}
Equating the second lines of \eqref{eqn:master_LHS} and \eqref{eqn:master_RHS} respectively, we find the following differential equation for the coherent states
\begin{align}
    \frac{\partial}{\partial t}\alpha_\mu &= 2\epsilon_\td(t)\text{sin}(\omega_\td t)e^{i\omega_\td t} -i(\Delta_\td+\tilde{\chi}_\mu)\alpha_\mu -\frac{\kappa}{2}\alpha_\mu \label{eqn:diff_A}
\end{align}
which reproduces \eqref{eqn:coherent_state_solution} in the main text upon substituting $\mu = jj$ and using the definition $\Delta_\td = \tilde{\omega}_\R - \omega_\td$ and \eqref{eqn:effective_chi}. Equating the last lines of \eqref{eqn:master_LHS} and \eqref{eqn:master_RHS} respectively gives the complex conjugate of \eqref{eqn:diff_A}.

We now find the solution to the above dynamical equation which is used in the next section to derive the SNR. Assuming the initial state of the resonator is the vacuum state, the solution is
\begin{equation}
    \alpha_\mu(t) = \int_0^tdt'\; 2\epsilon_\td(t')\text{sin}(\omega_\td t')e^{i\omega_\td t'}e^{-(\frac{\kappa}{2}+i(\Delta_\td+\tilde{\chi}_\mu))(t-t')}.
\end{equation}
Assuming the drive amplitude is constant $\epsilon_\td(t)\equiv \epsilon_\td$ the integral evaluates to
\begin{equation}
    \alpha_\mu(t) = \frac{-i\epsilon_\td}{\frac{\kappa}{2}+i(\tilde{\chi}_\mu+\Delta_\td)}\left(e^{-(\frac{\kappa}{2}+i(\tilde{\chi}_\mu+\Delta_\td))t}-1\right) +
    \frac{i\epsilon_\td e^{2i\omega_\td t}}{\frac{\kappa}{2}+i(\tilde{\chi}_\mu+\Delta_+)}\left(e^{-(\frac{\kappa}{2}+i(\tilde{\chi}_\mu+\Delta_+))t}-1\right)
\end{equation}
where $\Delta_\td=\tilde{\omega}_\R-\omega_\td$ and $\Delta_+=\tilde{\omega}_\R+\omega_\td$. When performing dispersive measurement, we will be driving on resonance such that $\omega_\td=\tilde{\omega}_\R$. In this case, we can drop the second term as $\Delta_+$ is large for the time scales and drive amplitudes we are considering. This is the rotating wave approximation applied to the drive term and the limits of this approximation are discussed in \cite{Zeuch_2020}. This approximation results in the simplified solution for the resonator coherent state
\begin{equation}\label{eqn:coherent_state}
    \alpha_\mu(t) = \frac{-i\epsilon_\td}{\frac{\kappa}{2}+i(\tilde{\chi}_\mu+\Delta_\td)}\left(e^{-(\frac{\kappa}{2}+i(\tilde{\chi}_\mu+\Delta_\td))t}-1\right).
\end{equation}
Note that $\alpha_\mu(0)=0$ which is consistent with the resonator initialised in the vacuum state.

For completeness, we finally find the solution for the coherences and amplitudes of the full quantum state $C_{\mu\nu}(t)$. To find $C_{\mu\nu}(t)$ we equate the first terms of \eqref{eqn:master_RHS} and \eqref{eqn:master_LHS}. By rewriting $2\epsilon_\td(t)\text{sin}(\omega_\td t)(e^{-i\omega_\td t}\alpha_\mu+e^{i\omega_\td t}\alpha_\nu^*)$ using equation \eqref{eqn:diff_A} we find
\begin{equation}
    \left(-\frac{1}{2}\frac{\partial}{\partial t}\left(|\alpha_\mu|^2\right)-\frac{1}{2}\frac{\partial}{\partial t}\left(|\alpha_\nu|^2\right)+\frac{\frac{\partial}{\partial t}C_{\mu\nu}}{C_{\mu\nu}}\right) = i(\tilde{\chi}_\nu-\tilde{\chi}_\mu)\alpha_\mu\alpha_\nu^*-\frac{\partial}{\partial t}(\alpha_\mu\alpha_\nu^*).
\end{equation}
We rearrange this to find a differential equation for the amplitude
\begin{equation}
    \frac{\partial}{\partial t}C_{\mu\nu} = \left(i(\tilde{\chi}_\nu-\tilde{\chi}_\mu)\alpha_\mu\alpha_\nu^*-\frac{\partial}{\partial t}\text{ln}(\braket{\alpha_\nu}{\alpha_\mu})\right)C_{\mu\nu}
\end{equation}
where we have used $\braket{\alpha_\nu}{\alpha_\mu}=\text{exp}(-\frac{1}{2}|\alpha_\mu|^2-\frac{1}{2}|\alpha_\nu|^2 +\alpha_\mu\alpha_\nu^*)$. The solution to this equation is 
\begin{equation}\label{eqn:amplitude}
    C_{\mu\nu}(t) = \frac{1}{\braket{\alpha_\nu(t)}{\alpha_\mu(t)}}e^{i(\tilde{\chi}_\nu-\tilde{\chi}_\mu)\int_0^t\alpha_\mu(t')\alpha_\nu^*(t')dt'}C_{\mu\nu}(0).
\end{equation}
Taking \eqref{eqn:trial_state}, \eqref{eqn:coherent_state} and \eqref{eqn:amplitude} together and given the initial condition $C_{\mu\nu}(0)$ we have an approximate solution to the Master equation \eqref{eqn:master_equation_S}. 

\subsection{D - The SNR for Homodyne Measurement}
In this section we discuss the theory of homodyne measurement as applied to the Master equation \eqref{eqn:master_equation_S}. This allows us to find the signal to noise ratio (SNR) for the dispersive readout of the TTR system \eqref{eqn:SNR_S}. 

As discussed in the main text, the transmons are measured by probing the resonator frequency using homodyne measurement. Homodyne measurement is performed by driving the resonator at its dressed frequency $\tilde{\omega}_\R$ and then mixing the reflected signal with a local oscillator (LO), also at frequency $\tilde{\omega}_\R$. The term $(\chi_1\sigma_{Z,1}+\chi_2\sigma_{Z,2})a^\dagger a$ in \eqref{eqn:H_disp_2_S} is a coupling between the transmon states and the resonator frequency. This shifts the resonator frequency, dependent on the state of the transmons $\ket{j_1j_2}$, to $\tilde{\omega}_\R-(-1)^{j_1}\chi_1-(-1)^{j_2}\chi_2$, $j_q\in\{0,1\}$, which, in turn, causes the phase of the reflected signal to shift. By measuring the phase difference between the LO and the reflected signal this shift can be determined \cite{Blais_2021, Krantz_2019}. 

Homodyne measurement outputs an integrated voltage signal $V_{\ket{j_1j_2},\tau}$, where $\tau$ is the measurement duration and $V_{\ket{j_1j_2},\tau}$ is a random variable with distribution $\mathcal{P}_{\ket{j_1j_2},\tau}(v)$. By inferring which distribution $V_{\ket{j_1j_2},\tau}$ came from using maximum likelihood estimation, the state of the transmons is measured \cite{Lalumiere_2010, Blais_2007, Filipp_2009}. 

We quantify the distinguishability between two integrated voltage signal distributions $\mathcal{P}_1$ and $\mathcal{P}_2$ with the SNR, defined  by \eqref{eqn:snr} in the main text, reproduced here as
\begin{equation}\label{eqn:snr_main_text_S}
    S\left(\mathcal{P}_1, \mathcal{P}_2\right) = \frac{2\left|\expV{V}{\mathcal{P}_1}-\expV{V}{\mathcal{P}_2}\right|}{\stdV{V}{\mathcal{P}_1} + \stdV{V}{\mathcal{P}_2}}
\end{equation}
where $\expV{V}{\mathcal{P}_1}$ ($\expV{V}{\mathcal{P}_2}$) is the mean integrated voltage and $\stdV{V}{\mathcal{P}_1}$ ($\stdV{V}{\mathcal{P}_2}$) the standard deviation.

The mean and standard deviation can either be calculated using input-output theory \cite{Blais_2021, Didier_2015} or the theory of continuous measurement \cite{Srinivas_1981, Wiseman_1993, Landi_2024}. Both theories give consistent results. By applying either of these theories to the dispersive Master equation \eqref{eqn:lindblad_disp_S}, it can be shown that the mean integrated voltage signal, produced by homodyne measurement of a transmon in the state $\ket{\mu}$, is 
\begin{equation}\label{eqn:e_diff}
    \expV{V}{\mathcal{P}_{\ket{\mu},\tau}} =\sqrt{\eta\kappa z}\int_0^\tau(e^{-i\phi}\alpha_\mu(t)+e^{i\phi}\alpha_\mu^*(t))dt.
\end{equation}
and the standard deviation is
\begin{equation}\label{eqn:var_diff}
    \stdV{V}{\mathcal{P}_{\ket{\mu},\tau}}=\sqrt{z\tau}
\end{equation}
where $\eta$ is the detector efficiency, $\phi$ is the phase difference between the local oscillator and the reflected signal, $\alpha_\mu(t)$ is the coherent state of the resonator as defined in \eqref{eqn:coherent_state}, $\tau$ is the measurement duration and $z$ is the decay rate of the local oscillator. $z$ does not feature in the final result but we include it to make the mean and standard deviation dimensionless.

Taking $\mathcal{P}_1=\mathcal{P}_{\ket{\mu},\tau}$ and $\mathcal{P}_2=\mathcal{P}_{\ket{\nu},\tau}$ and inserting \eqref{eqn:e_diff} and \eqref{eqn:var_diff} into \eqref{eqn:snr_main_text_S} we find 
\begin{equation}\label{eqn:SNR_S}
    S(\tau) = 2\sqrt{\frac{\eta\kappa}{\tau}}\left|\int_0^\tau\text{Re}[(\alpha_\mu(t)-\alpha_\nu(t))e^{-i\phi}]dt\right|.
\end{equation}
With $\mu=00$, $\nu=11$ and $\phi=0$ this reproduces \eqref{eqn:snr_continuous_measurement} in the main text.

Note that this equation also applies to the TR system discussed in SM A where the coherent state of the resonator is the same as in \eqref{eqn:coherent_state} but with $\tilde{\chi}_\mu = \bra{\mu}\chi\sigma_{Z}\ket{\mu}$ where $\chi$ is as defined in \eqref{eqn:H_disp_s}.

We insert our solution for the coherent state \eqref{eqn:coherent_state} into \eqref{eqn:SNR_S} and, using $\Delta_\td=0$ since we drive on resonance for dispersive measurement, we find 
\begin{align}
    S(\tau) &= 4\epsilon_\td|\text{cos}(\phi)|\sqrt{\frac{\eta\kappa}{\tau}} \left| \frac{(\chi_1+\chi_2) \tau }{\frac{\kappa^2}{4}+(\chi_1+\chi_2)^2} \right.\nonumber \\
    &\quad\left.-\frac{\kappa(\chi_1+\chi_2)(1-e^{-\frac{\kappa}{2}\tau}\text{cos}((\chi_1+\chi_2)\tau))}{(\frac{\kappa^2}{4}+(\chi_1+\chi_2)^2)^2} +\frac{(\frac{\kappa^2}{4}-(\chi_1+\chi_2)^2)e^{-\frac{\kappa}{2}\tau}\text{sin}((\chi_1+\chi_2) \tau)}{(\frac{\kappa^2}{4}+(\chi_1+\chi_2)^2)^2}\right|
\end{align}
where we have used $\tilde{\chi}_{00}=-\tilde{\chi}_{11}=-\chi_1-\chi_2$, which follows from \eqref{eqn:effective_chi}. In the steady state regime $\tau\gg\frac{2}{\kappa}$ the SNR is maximised when $\phi=0$ giving
\begin{equation}\label{eqn:snr_ss_S}
S_\tss(\tau)=\frac{4\epsilon_\td\sqrt{\eta\kappa}|\chi_1+\chi_2|}{\frac{\kappa^2}{4}+(\chi_1+\chi_2)^2}\sqrt{\tau}.
\end{equation}

The SNR could in principle be improved by increasing the drive amplitude $\epsilon_\td$ but this is limited by the number of photons in the resonator as we need to keep this photon number below the critical photon number, as discussed in the main text. The number of photons in the resonator in the steady state $n_{\tss}$ is found by taking \eqref{eqn:coherent_state} to the steady state regime
\begin{equation}
 n_{\tss}=\lim_{t\gg\frac{2}{\kappa}}|\alpha_{00}(t)|^2=\lim_{t\gg\frac{2}{\kappa}}|\alpha_{11}(t)|^2=\frac{\epsilon_\td^2}{\frac{\kappa^2}{4}+(\chi_1+\chi_2)^2}.
\end{equation}
We rewrite $\epsilon_\td$ in terms of $n_{\tss}$, as this is the quantity we want to keep fixed, and we find
\begin{equation}\label{eqn:snr_ss1_S}
    S_\tss(\tau)=4\sqrt{\eta n_{\tss}}\sqrt{\frac{\kappa(\chi_1+\chi_2)^2}{\frac{\kappa^2}{4}+(\chi_1+\chi_2)^2}}\sqrt{\tau}.
\end{equation}
Keeping all other parameters constant, this function is maximised for $\kappa=2|\chi_1+\chi_2|$ which when re-substituted back into \eqref{eqn:snr_ss1_S} gives \eqref{eqn:snr2_ss} from the main text.

\subsection{E - Semi-Classical Branch Analysis for the Double Transmon System}
In this section we extend the semi-classical Floquet branch analysis method developed in \cite{Dumas_2024}, used for finding the critical photon number of a TR system, to the TTR system. We do this as performing full branch analysis on the TTR system, as described in Appendix C, is numerically intractable. The semi-classical approach allows us to trace out the resonator subspace while maintaining an effective drive on the transmons, significantly reducing the computational resources required to perform branch analysis on our system.

In contrast to the analysis performed in sections B, C and D we do not make any approximations on the transmon subspace of the Hamiltonian \eqref{eqn:htr_S}-\eqref{eqn:coupling_term_2} beyond dropping $n_{\g,q}$ from $V_q$ \eqref{eqn:coupling_term_2}, replacing $V_q$ with $V'_q=-ig_{q} n_{\T,q} (a-a^\dagger)$, as in \eqref{eqn:simplified_coupling_TTR}-\eqref{eqn:2_tmon_couple}. We also truncate both transmons to 18 energy levels which allows us to numerically study the full transmon dynamics, as the resonator is driven, in order to find the critical photon number. Consider the Master equation for the full system before the dispersive approximation has been made
\begin{equation}
    \frac{\partial}{\partial t}\rho = -i[H'_{\TTR} + H_{\td}(t),\rho] + \kappa\mathcal{D}[a]\rho.
\end{equation}
where $H'_{\TTR}$ is defined in \eqref{eqn:simplified_coupling_TTR}-\eqref{eqn:2_tmon_couple}, $H_\td(t)$ is defined in \eqref{eqn:drive_S} and $\mathcal{D}[a]\rho = a\rho a^\dagger - \{a^\dagger a, \rho\}/2$ is the Lindblad dissipator. In order to trace out the resonator subspace we first transform the Master equation using the displacement operator $D(\beta(t))=\text{exp}(\beta(t) a^\dag - \beta(t)^* a)$ such that we are in a frame which follows the resonator state throughout its evolution. In this frame the resonator drive is eliminated and an effective drive is induced on the transmon. The Master equation in the displaced frame is
\begin{align}\label{eqn:displaced_master}
    \frac{\partial}{\partial t}\rho_D &= -i[H'_{\TTR} - 2\sum_{q=1}^2g_q\text{Im}(\beta)n_{\T,q}, \rho_D] + \kappa D[a]\rho_D \nonumber \\ 
    &\quad + [a,\rho_D](-\frac{\partial}{\partial t}\beta^*-2\epsilon_\td(t)\text{sin}(\omega_\td t)+i\omega_\R\beta^*-\kappa\frac{\beta^*}{2}) \nonumber \\
    &\quad + [a^\dag,\rho_D](\frac{\partial}{\partial t}\beta+2\epsilon_\td(t)\text{sin}(\omega_\td t)+i\omega_\R\beta+\kappa\frac{\beta}{2})
\end{align}
where $\rho_D=D\rho D^\dag$. If we choose $\beta(t)$ such that it satisfies the dynamical equation 
\begin{equation}\label{eqn:displaced_dynamical}
\frac{\partial}{\partial t}\beta=-2\epsilon_\td(t)\text{sin}(\omega_\td t)-i\omega_\R\beta-\kappa\frac{\beta}{2}
\end{equation}
the drive on the resonator is eliminated and we induce an effective drive $- 2\sum_{q=1}^2g_q\text{Im}(\beta)n_{\T,q}$ on the two transmons. The resonator is effectively displaced to the vacuum state. Note, this equation is very similar to \eqref{eqn:diff_A} except it does not include the dispersive shift due to the transmon state.

At this point in the calculation for the TR system the resonator subspace would be traced out \cite{Dumas_2024}, however, this neglects quantum fluctuations in the resonator. These fluctuations are inconsequential for the TR system but for the TTR system they can mediate resonances between the two transmons which can result in an abrupt transfer of population between them, a physical effect not present in the TR system. Before tracing out the resonator, we must therefore approximate this transmon-transmon exchange by performing a SWT on the displaced Hamiltonian. This will retain the second order coupling between the transmons induced by the resonator.

We perform the SWT $e^S$ on \eqref{eqn:displaced_master} using the same transformation operator $S$ as defined in \eqref{eqn:SWT_operator}. We have already calculated the SWT of the bare Hamiltonian $H'_{\TTR}$ in \eqref{eqn:SWT-expansion-2}. As in the previous analysis, we drop the higher order terms $\mathcal{O}(g^3)$ but in this analysis we keep all the off diagonal terms which were previously neglected. Previously, dropping the higher order terms $\mathcal{O}(g^3)$ neglected important physics which became relevant as the resonator photon number increased however, in this case, we have first displaced the resonator to the vacuum state so these higher order terms never become relevant. The important physics we want to capture at large resonator photon numbers is instead captured by the effective transmon drive. 

The SWT applied to the transmon drive $-2\sum_{q=1}^2g_q\text{Im}(\beta)n_{\T,q}$, once higher order terms have been dropped, only includes terms which act on the resonator which will be traced out when we make the semi-classical approximation. The SWT applied to the decay term results in Purcell decay which we neglect as we assume a Purcell filter is used, as discussed in SM B. 

We now make the semi-classical approximation by tracing out the resonator, dropping all terms acting non-trivially on the resonator subspace. The semi-classical dynamics of the two transmons is therefore fully described by the Floquet Hamiltonian
\begin{equation}\label{eqn:floquet}
    H_F(t) = \sum_{q=1}^2\sum_{j} \left(E_{j,q} + \Lambda_{jj,q}\right)\ket{j_q}\bra{j_q} +  \sum_{q=1}^2\sum_{\substack{i,j \\ i\neq j}} \Lambda_{ij,q}\ket{i_q}\bra{j_q} + \frac{1}{2}\sum_{\substack{i,j\\k,l}} J_{ijkl}\ket{i_1k_2}\bra{j_1l_2} + \sum_{q=1}^2\tilde{\epsilon}_q(t)n_{\T,q}
\end{equation}
where the symbols are as defined in \eqref{eqn:SWT-expansion-2} and $\tilde{\epsilon}_q(t)=-2g_q\text{Im}(\beta(t))$.

Given the initial vacuum state, the solution to \eqref{eqn:displaced_dynamical} is $\beta(t)=-2\int_0^tdt'\epsilon_\td(t')\text{sin}(\omega_\td t')e^{-(i\omega_\R+\frac{\kappa}{2})(t-t')}$. Choosing $\omega_\td=\omega_\R$ and a constant drive $\epsilon_\td(t')=\epsilon_\td$ we find
\begin{equation}\label{eqn:alpha_s}
    \beta(t) = -\epsilon_\td\left(\frac{e^{-i\omega_\R t}-e^{-(i\omega_\R+\frac{\kappa}{2})t}}{-i\kappa/2}-\frac{e^{i\omega_\R t}-e^{-(i\omega_\R+\frac{\kappa}{2})t}}{2\omega_\R-i\kappa/2}\right)
\end{equation}
and therefore 
\begin{equation}\label{eqn:floquet_drive}
    \tilde{\epsilon}_q(t)=2g_q\sqrt{n(t)}\text{cos}(\omega_\R t)
\end{equation}
where we have neglected the second term of \eqref{eqn:alpha_s} since $\omega_\R\gg \epsilon_\td$ and we have used $\sqrt{n(t)}=|\beta(t)|\approx\frac{2\epsilon_\td}{\kappa}(1-e^{-\frac{\kappa}{2}t})$ where $n(t)$ corresponds to the photon number of the resonator at time $t$.

Using \eqref{eqn:floquet} and \eqref{eqn:floquet_drive}, Floquet branch analysis is performed as described in \cite{Dumas_2024} in order to extract the critical photon number for the TTR system. This is done across a range of parameters and the results are summarised in Fig. \ref{fig:two_tmon_n_crit_violations} in the main text.

\subsection{F - Code}
All the code used to generate the data and plots in this paper can be found at the GitHub link: \url{https://github.com/custal/ancilla-assisted-accelerated-readout-of-transmon-qubits}

\end{document}